\documentstyle[12pt]{article}
\def\beq{\begin{equation}}
\def\eeq{\end{equation}}
\def\bea{\begin{eqnarray}}
\def\eea{\end{eqnarray}}
\def\bq{\begin{quote}}
\def\eq{\end{quote}}
\def\bi{\begin{itemize}}
\def\ei{\end{itemize}}
\def\beqa{\begin{eqnarray}}
\def\eeqa{\end{eqnarray}}
\def\be{\begin{enumerate}}
\def\ee{\end{enumerate}}
\def\beq{\begin{equation}}
\def\eeq{\end{equation}}
\def\bi{\begin{itemize}}
\def\ei{\end{itemize}}

\def\pa{\partial}

\def\cp{{\cal P}}
\def\cl{{\cal L}}

\def\cq{{\cal Q}}
\def\cR{{\cal R}}
\def\cS{{\cal S}}

\def\r2{\sqrt{2}}
\def\ra{\rightarrow}
\def\bi{\begin{itemize}}
\def\ei{\end{itemize}}

\def\ov{\overline}
\def\nn{\nonumber \\}

\def\ca{{\cal A}}

\def\be{\begin{enumerate}}
\def\ee{\end{enumerate}}

\def\bc{\begin{center}}
\def\ec{\end{center}}

\def\pa{\partial}

\def\ta{\tilde a}

\def\cp{{\cal P}}
\def\cl{{\cal L}}

\def\r2{\sqrt{2}} \def\rt{\sqrt{2}}
\def\ra{\rightarrow}

\def\ov{\overline}
\def\nn{\nonumber \\}

\def\ca{{\cal A}}

\begin{document}
\pagestyle{empty}
\begin{flushright}   IFT/2002/25, T/02-089
\end{flushright}
\vskip 2cm
\begin{center}
{\Huge Brane world supersymmetry, detuning, flipping and
orbifolding}
\vspace*{5mm} \vspace*{1cm}
\end{center}
\vspace*{5mm} \noindent
\vskip 0.5cm
\centerline{\bf Philippe Brax${}^{1}$ and Zygmunt Lalak${}^{2}$}
\vskip 1cm
\vskip 0.3cm
\centerline{\em ${}^{1}$ Service de Physique Th\'eorique}
\centerline{\em CEA-Saclay F-91191 Gif/Yvette, France}
\vskip 0.3cm
\centerline{\em ${}^{2}$Institute of Theoretical Physics}
\centerline{\em University of Warsaw, Poland}
\vskip 2cm

\centerline{\bf Abstract}
\vspace{1.0cm}
\noindent We emphasize the necessity of a delicate interplay between
the  gauge
and gravitational sectors of five-dimensional brane worlds in creating phenomenologically relevant vacua. We discuss  locally supersymmetric brane worlds with unflipped and flipped
fermionic boundary conditions and with matter on the branes.
We point out that a natural separation between the gauge and gravity sectors,
very difficult in models with true extra dimensions, may be achieved in 4d models
with deconstructed dimensions.
\vskip .3cm
\vskip4cm
\newpage
\pagestyle{plain}
\section{Introduction}
There exist locally supersymmetric
theories in five dimensions that include nontrivial physics localized on
four-dimensional branes \cite{bagger,gp,flp,flp2,kallosh,bfl,Bagger:2001qi,Gherghetta:2002nr,Bagger:2002rw}. The brane sectors may contain arbitrary
four-dimensional gauge theories as well as localized interactions between
bulk fields. In the bulk one has a gauged 5d supergravity, which can be
further coupled to five-dimensional gauge sectors. Such five-dimensional
models with branes are likely to lead to interesting extensions of
the Standard Model, pertaining to novel approaches to the hierarchy problem.
In general it is easier to separate (super)gravity from fields which are charged under the gauge group,
and to study  gobally supersymmetric five-dimensional gauge models with branes, instead of
working with the complete matter-supergravity Lagrangian. However, in the context of  brane worlds,
where the geometry of the extra dimension may play a nontrivial role, this approach is not sufficient.
The potentials that appear in the bulk and on the branes due to gauge sector interactions
couple to moduli fields  which serve as sources in the Einstein equations, hence they back-react on
the geometry. In fact, the gauge sector potential should be studied simultaneously with the issue of the stability
of the orbifold, otherwise the conclusions drawn from the simplified models with decoupled gravity may turn out to be misleading.\\
The issue of constructing an explicit model with a general bulk and brane nonabelian gauge sectors
coupled consistently to supergravity is a fairly complex one, and at present one has to rely on simpler
constructions. In this paper we summarize the attempts made in this direction in published work, and in forthcoming publications.

\section{Five-dimensional supergravities on $S^1/Z_2$}
\subsection{Supergravity Lagrangian: detuning and flipping}
Let us summarize  the basic features of pure 5d N=2 gauged supergravity on $S^1/Z_2$. The gravity multiplet $(e_\alpha^m, \psi_\alpha^A, \ca_\alpha)$ consists of  the vielbein, a pair of symplectic Majorana gravitini, and a vector field called the graviphoton. There is a global SU(2) R-symmetry which rotates the two supercharges into each other. Making use of the graviphoton we can gauge a U(1) subgroup of the R-symmetry group. Such gauging can be described by an SU(2) algebra valued matrix $\cp=\vec{P} \cdot  i\vec{\sigma}$ (prepotential). We do not give the complete form of the  action and supersymmetry transformation laws in gauged supergravity (see \cite{agata} for details), but only the relevant
terms. The gravitino transformation law gets the following correction due to gauging (we use the normalization of \cite{ovrut})
\beq
\label{susygrav}
\delta \psi_\alpha^A =
-i\frac{\r2}{3}\gamma_\alpha g\cp^A_B \epsilon^B
\eeq
where $g$ is the U(1) gauge charge. Gauging introduces also the potential term into the action
\beq
V =\frac{8}{3} g^2 Tr (\cp^2).
\eeq
Without bulk matter fields the prepotential is just a constant matrix so the potential term corresponds to a (negative) cosmological constant.

The distinguishing feature of the brane-bulk scenario is that the fifth dimension is an orbifold $S_1/Z_2$
with branes located at the fixed points. It is equivalent  to work on a smooth circle $S_1$ with the fifth coordinate ranging from $-\pi\rho$ to $\pi\rho$ and impose $Z_2$ symmetry on the fields of the Lagrangian.
The $Z_2$ symmetry acts by $x^5 \ra -x^5$. Under its action the bosonic fields $g_{\mu\nu}$, $g_{55}$ and  $\ca_5$ are even, while $g_{\mu 5}$ and $\ca_\mu$ are odd. The $Z_2$ action on the gravitino is defined as follows
\bea
\label{z2}
\psi^A_\mu(-x^5)= \gamma_5 \cq^A_B \psi_\mu^B(x^5) & \psi^A_5(-x^5)= -\gamma_5 \cq^A_B \psi_5^B(x^5)
\eea
where $\cq=\vec{Q} \cdot  \vec{\sigma}$ and $\vec{Q}^2=1$.
The $Z_2$ action on the supersymmetry generating parameter $\epsilon$ must be the same as that on the 4d components of the gravitino.

Further we need to define the $Z_2$ symmetry under reflection around the second fixed point at $x^5=\pi\rho$
\bea
\label{z2f}
\psi^A_\mu(\pi\rho-x^5)= \alpha \gamma_5 \cq^A_B \psi_\mu^B(\pi\rho+x^5) & \psi^A_5(\pi\rho-x^5)= -\alpha\gamma_5 \cq^A_B \psi_5^B(\pi\rho+x^5).
\eea
Apart from the conventional case $\alpha=1$, in this letter we also consider the `flipped' supersymmetry  with $\alpha=-1$.  In the latter case supersymmetry is always broken globally, as different spinors survive the orbifold projection on each wall. Note also, that in the flipped case we have $\psi_\alpha^A(x^5+2\pi\rho)=-\psi_\alpha^A(x^5)$.

It is straightforward to check that the  5d {\it ungauged}
supergravity action  is invariant under transformations (\ref{z2}) and
(\ref{z2f})  but the {\it gauged} supergravity action  is not invariant if the prepotential $\cp$ is a general one.  The action and the supersymmetry transformation laws are  $Z_2$ invariant if we choose the prepotential in the form:
\beq
\label{ourpre}
g\cp = g_1 \epsilon(x^5) \cR +  g_2 \cS
\eeq
where $\cR=\vec{R} \cdot i\vec{\sigma}$ commutes with $\cq$ and $\cS=\vec{S} \cdot i\vec{\sigma}$ anticommutes with $\cq$. Equivalently,  $\cR= i \sqrt{\vec{R}^2} \cq$ and  $\cS = (\vec{Q}\times \vec{U}) \cdot i\vec{\sigma}$ with some arbitrary vector $\vec{U}$. Note that the cosmological constant  does not contain the step function and is given by
\begin{equation}
\Lambda_5=-\frac{16}{3}(g_1^2\vec{R}^2 + g_2^2\vec{S}^2).
\end{equation}
The supergravity action with prepotential (\ref{ourpre}) contains both symmetric  and antisymmetric (multiplied by $\epsilon(x^5)$) gravitino masses.
The important thing to note is that
the presence of the $Z_2$-symmetric piece $\cS$ in the prepotential results in
the supersymmetric detuning between the brane tensions and the bulk cosmological term.
This detuning takes place independently of the value of the `flip' parameter $\alpha$.
If $\cS$ is set to zero, and $\alpha=+1$, then the supersymmetric relation between brane and bulk tensions results in
a warp factor that is precisely the one of the Randall-Sundrum model.

The part of the gravitino transformation law due to gauging  is now:
\beq
\label{susygrav1}
\delta \psi_\alpha^A =
-i \frac{\r2}{3}\gamma_\alpha (g_1\epsilon(x^5)\cR^A_B+g_2 \cS^A_B)\epsilon^B.   \eeq
The presence of the step function in the above transformation law
implies that the 5d action is not supersymmetric. The fifth derivative
in the gravitino kinetic term acts on the step function producing an
expression multiplied by a  delta function. The uncancelled variation is:
\bea
\label{varo}
\delta \cl = -2i\sqrt{2}g_1(\delta(x^5)-\delta(x^5-\pi\rho)) e_4 \cR^A_B \ov{\psi_\mu}_A \gamma^\mu \gamma^5 \epsilon^{B}.
\eea
Notice that when $g_1=0$ the above variation  vanishes
implying that the Lagrangian is supersymmetric.
Using the fact that  the matrix  $\cR$ is proportional to $\cq$ we have the following relations:
\bea
\gamma_5\cR^A_B \epsilon^B(0)= i\sqrt{\vec{R}^2}\epsilon^A(0)
\nn
\gamma_5\cR^A_B \epsilon^B(\pi\rho)= i\alpha\sqrt{\vec{R}^2}\epsilon^A(\pi\rho).
\eea
Thus:
\bea
\label{var}
\delta \cl = 2\sqrt{2}g_1\sqrt{\vec{R}^2} e_4\ov{\psi_\mu}_A \gamma^\mu \epsilon^{A} (\delta(x^5)-\alpha\delta(x^5-\pi\rho)).
\eea
The variation (\ref{var}) can be cancelled by the variation of the determinant in the brane tension term:
\beq
\cl_T= -4 \sqrt{2}g_1\sqrt{\vec{R}^2} e_4 (\delta(x^5)-\alpha\delta (x^5-\pi\rho)).
\eeq
Summarizing (and changing the normalization to that used by Randall and Sundrum), we constructed a {\em locally  supersymmetric Lagrangian}, which has the following bosonic gravity part:
\beq
M^{-3} S= \int d^5 x \sqrt{-g_5} (\frac{1}{2}R + 6 k^2)-  6 \int d^5 x\sqrt{-g_4}k T (\delta(x^5) - \alpha \delta(x^5-\pi\rho))
\eeq
where we have defined
\begin{equation}
k = \sqrt{\frac{8}{9}(g_1^2 R^2+ g_2^2S^2)}
\end{equation}
 and
\begin{equation}
T= \frac{g_1\sqrt{\vec{R}^2}}{\sqrt{(g_1^2 R^2+ g_2^2S^2)}}.
\end{equation}

The BPS relation between the bulk cosmological constant and the brane
tensions (`the Randall-Sundrum fine-tuning') corresponds to $T=1$,
which holds only when $\langle \cS \rangle=0$. In such case the vacuum
solution is  $AdS_5$ in the bulk  with flat Minkowski branes
\cite{rs1}. This vacuum preserves one half of the supercharges
corresponding to unbroken N=1 supersymmetry in four dimensions
\cite{flp}. As soon as we switch on non-zero $\cS$, we get $T < 1$,
the BPS relation  is destroyed and the vacuum breaks
all supersymmetries.

The $N=1$ supersymmetry is broken when $\alpha=\pm 1$ and $\langle \cS \rangle
\neq 0$.
If $\alpha =-1$, then supersymmetry is always broken globally,
independently of the expectation value of $\cS$. To see this explicitly,
let us take $\langle \cS \rangle = 0$ and
note that the fermions which are allowed to propagate on the left
and right branes have to obey the conditions $W^{A}_{0\;B} \psi^B=0$ and
$W^{A}_{\pi \;B} \psi^B=0$ respectively, where $W_{0,\;\pi}$ are given by
(\ref{z2f}). The projection operators $\Pi^{A}_{\pm \;B} = \frac{1}{2}
( {\bf 1}  \delta^{A}_B \pm \gamma_5 \cq^{A}_B )$ split each spinor
into two components, one of which is annihilated by $W_0$:
$W_0 \epsilon_{+} = W_0 \Pi_{+} \epsilon = 0$. The second component, $\Pi_{-}
\epsilon$,
is annihilated by $W_\pi$ if $\alpha=-1$; for $\alpha=+1$ $W_0 = W_\pi$.
The BPS conditions imply (we take here $ds^2 = a^2 (x^5) dx^2 + (d x^5)^2 $)
\beq \label{bps3} \frac{a'}{a} \gamma_5 \epsilon^A + \frac{2 \sqrt{2}}{3} g_1 \epsilon(x^5)
\sqrt{R^2} \cq^{A}_B \epsilon^B = 0 \eeq
(this holds for $AdS_4$ and Minkowski foliations).
When we apply the operator $\Pi_{+}$ to (\ref{bps3}), we obtain the conditions
$\frac{a'}{a} + \frac{2 \sqrt{2}}{3} g_1 \epsilon(x^5)
\sqrt{R^2}=0$ or $\epsilon_{+} \equiv 0$. The first possibility
leads to discontinuities of the warp factor at the fixed points:
$ [ \frac{a'}{a} ]_0 = - 2 k T$, $[ \frac{a'}{a} ]_{\pi \rho} = + 2 k T$.
However, the matching conditions in the equations of motion give
$ [ \frac{a'}{a} ]_0 = - 2 k T$, $[ \frac{a'}{a} ]_{\pi \rho} = + 2
\alpha k T$, which are in contradiction with the BPS condition for
$\alpha =-1$, unless $\epsilon_{+} \equiv 0$. Applying to (\ref{bps3}) the
second projector, $\Pi_{-}$, one finds out immediately that boundary
conditions and BPS conditions agree on both branes only
for $\epsilon_{-} \equiv 0$.
Thus there exists no globally defined Killing spinor in the setup with
flipped $Z_2$ acting on fermions (bosons are acted on as in the unflipped case), and all supersymmetries are broken spontaneously.

Now we move on to the `flipped susy'  case $\alpha=-1$. A vacuum solution  in the warped product form can be found
\begin{equation}
ds^2 = a^2(x^5)g_{\mu\nu}dx^\mu dx^\nu + R_0^2 (dx^5)^2,
\end{equation}
where $g_{\mu\nu}$ is the $AdS_4$ metric with cosmological constant $\bar \Lambda$
\beq
g_{\mu\nu}dx^{\mu}dx^{\nu}=e^{-2\sqrt{-\bar\Lambda}x_3}(-dt^2+dx_1^2+dx_2^2) +dx_3^2
\eeq
and the third coordinate $x_3$ has been singled out.
As long as $T<1$  the static vacuum solution is $AdS_5$ in the bulk  and  the warp factor  can be parametrized as \cite{dewolfe,rk}:
\beq
\label{ads4sol}
a(x^5) = \frac{\sqrt{-\bar \Lambda}}{k}\cosh (k R_0 |x_5| - C).
\eeq
The matching  conditions  for $AdS_4$ branes embedded in $AdS_5$ read
\bea
\tanh(C)&=&  T \\
\tanh (k R_0 \pi\rho - C)&=& -\alpha  T\; = \;  T .
\eea
The first condition sets the integration constant $C$ and the second fixes the size of the fifth dimension.
The radion is stabilized at the value
\begin{equation} \label{radfix}
\pi\rho k R_0= \ln(\frac{1+T}{1-T}).
\end{equation}
Moreover, the magnitude of the brane cosmological constant is fixed by
the normalization $a(0)=1$. This leads to
\begin{equation}
\bar \Lambda=(T^2 - 1)k^2 < 0.
\end{equation}
The cosmological constant on the brane depends directly on the scale of supersymmetry breaking on the brane. The same is true for
the expectation value of the radion. Notice, that $\langle R_0 \rangle$
can be expressed solely in terms of $k$ and $\bar \Lambda$.
The formula equivalent to (\ref{radfix}) in the case $\alpha=+1$
gives $R_0 = 0$.
To summarize, the nonzero expectation value of $\cS$ gives rise to detuning
between brane and bulk tensions and as a consequence to stabilization of
the radion. The $\langle \cS \rangle $ contributes also to supersymmetry breakdown, but in the case of $\alpha=-1$ supersymmetry is broken even if
$\langle \cS \rangle = 0$.

One finds that (\ref{ads4sol}) is not a valid solution in the case $T=1$.
Indeed, this implies that the brane cosmological
constant vanishes and the second brane is sent to infinity.
In that case we expect a global mismatch due to the boundary
condition on one of
the branes. It turns out that a static solution with maximally symmetric foliations do not exist in this case, but one can find cosmological solutions which may be considered  as a background
for the physics on the brane \cite{bfl}.

\subsection{Matter on the branes and in the bulk: backreaction on geometry}
After describing the supergravity background for brane world models, one needs to enhance them
by putting matter, scalar  and gauge fields on the branes and in the bulk.
Although some work on bulk gauge theories coupled to supergravity is
available, in order   to
perform a detailled analysis it  is convenient to restrict the standard gauge sectors to the branes,
and to leave in the bulk only matter supermultiplets coupled to the particular local $U(1)_R$,
which is the gauged subgroup of the R-symmetry of the $N=2$ supersymmetry algebra.
This role is conveniently played by the bulk universal hypermultiplet (the gauge field being the graviphoton), whose introduction is also well motivated by stringy considerations.
Nonabelian gauge sectors on the branes are both models for the Standard Model physics, and simultaneously  supply the seeds for the supersymmetry breakdown. The situation in the $\cS=0$, $\alpha=+1$ case has been discussed in great detail in \cite{flp3}. \\
The signature  of  supersymmetry breakdown in these models is the nonzero
expectation value of the $Z_2$-odd complex scalar from the hypermultiplet, $\xi$, and of its transverse derivative $\partial_5 \xi$. To excite a nontrivial vacuum configuration for this field one needs to
switch on its sources on the branes. These sources can be represented by the effective superpotentials on the branes,
 $W_i, \, i=1,2$. However, once this is done, new contributions to energy densities on the branes and in
the bulk are created. This leads to modifications of the vacuum configuration of the moduli and of the warp factor. In the case where the purely gravitational background had 4d flat foliations, the
backreaction of the supersymmetry breaking physics leads naturally towards anti-de Sitter geometry on the branes, with negative four-dimensional cosmological constant. One should notice, that the non-zero value of $\cS$
would not help, since the way it acts is to make the size of the bulk cosmological constant larger, but without changing its sign, so that it remains negative.
The good thing that happens is that the expectation value of the radion, hence the distance between
branes, becomes determined (it is a modulus as long as supersymmetry is preserved)  in terms of the supersymmetry breaking sources. However, to obtain
the required size of mass splitting within supermultiplets and the right hierarchy between the 4d Planck scale and the electroweak scale
one needs a tuning of the sources, which signals an instability.
To see this more explicitly consider the issue of making the effective four-dimensional cosmological constant zero by including positive contributions to the
potential. A first obvious source for such contributions  are the
F-terms borne  by  the matter sector localized on the branes. They contribute the terms
$
\delta V_{boundary} = \frac{1}{2 V} \delta (x^5) | \frac{\pa W_1}{\pa \Phi_1}
|^2 + \frac{1}{2 V} \delta (x^5 - \pi \rho) | \frac{\pa W_2}{\pa \Phi_2}
|^2$.\\
However, this modification does not work on its own and the 4d geometry stays anti-de Sitter.
To find a vacuum with  the flat geometry one needs to create a potential potential for the second bulk field, the dilaton S, and in addition to put a Polonyi field on the Planck brane.
The effective 4d superpotential that does the job is
\beq
W = W_1 (\Phi_1) + (e^{-a_1 S} + d e^{-a_2 S}) e^{-3 T}
\eeq
with $|W_1'|^2 \approx 2 W_{1}^2$.
One obtains in this case $V_0 \approx 1/a$, $e^{k \pi \rho R_0} \approx |W_2|/|W_1|$,
and $F^{\Phi_1}$ and $F^S$ become dominant. Mass splittings are due to
universal soft scalar masses $\sim m^{2}_{3/2}$, $A_3$ terms $\sim m_{3/2}$,
and gaugino masses $\sim m_{3/2}$, that are universal due to $F^S \gg F^T$.
The price one has to pay for the vanishing cosmological constant is the active role of the Polonyi field.
In general, this example amplifies the observation, that to obtain   a phenomenologically
interesting vacuum in brane models one needs certain correlations between parameters of the different  brane sectors. This is somewhat unnatural in view of the fact, that the branes are spatially
separated.

It is interesting to extend this discussion to the flipped case, where $\alpha=-1$.
The observations are likely to be relevant for the case of the stringy brane-antibrane models,
and for models similar to these of Barbieri, Hall and Nomura \cite{Barbieri:2001vh}.
The gravitational background for the flipped case has been discussed earlier, let us only remind here
that there is no static solution with the Minkowski foliation, but we have found solutions with
anti-de Sitter foliation and stable radion due to the introduction of the detuning parameter $\cS$.
Hence we should couple bulk and brane matter to such a background with negative cosmological constant in 4d, and try to cancel the
cosmological constant   dynamically by positive contributions coming from the branes.
It is easy to see that one can include the bulk universal
hypermultiplet in  the same way as
in the unflipped case. The general difference is that now on the flipped brane certain terms needed to compensate delta-type variations of the bulk terms will have the  opposite
sign to that on the
unflipped brane. The reason for that is precisely the same as the change of sign of the brane tension
on the flipped brane in the purely gravitational case.
The second important difference is the coupling of the bulk fermions to the flipped brane.
At this brane the fermions that couple to brane operators are the $Z_2$-odd components of the bulk
symplectic-Majorana fermions (gravitini and hyperini), while the components of these fermions that enter the unflipped brane are the $Z_2$-even ones. More precisely, the relevant parts of the flipped-brane-bulk
coupling are $S = S_{bulk} + S_{YM} + S_{matter}$ where
\beqa \label{eq:acc}
&S_{bulk} = \int d^5 x \; e_5 (\frac{R}{2} - \frac{1}{V} |\xi' |^2 g^{55}
-\frac{1}{4 V^2} V'^2 g^{55}+ \Lambda^2 (\frac{1}{6} + \frac{1}{12 V}
|\xi|^2 - \frac{1}{12 V^{2} } |\xi|^4 ))& \nonumber \\
&S_{YM}= \int d^5x \frac{e_5}{e_5^5}\delta(x^5)
\left ( \right .
 -\frac {V} {4}
F_{\mu\nu}^{a} F^{a\mu\nu}
-\frac{1}{4}
\sigma F_{\mu\nu}^{a} \tilde F^{a\mu\nu}
-\frac {V} {2}
\overline{\chi^{a}}D\!\!\!\!\slash\chi^{a}& \nonumber \\
&-\frac{\sqrt{V}}{2e_{5}^{5}}
(
(\overline{\chi^{a}}_{L}\chi^{a}_{R})\partial_{5}\overline{\xi}
+ (\overline{\chi^{a}}_{R}\chi^{a}_{L})\partial_{5}\xi
)
+ \delta(0)
\frac{V^{3/2}}{8} ((\chi^{a})^2)^2 )
+ ...
\left .   \right )& \nonumber \\
&S_{i \;matter}= \int d^5 x \; \frac{e_5}{e^{5}_{5}}  \delta(x^5- x_i ) ( -
\epsilon_i\Lambda
(1-\frac{|\xi|^2}{V})
  -  \frac{2}{V} \sqrt{g^{55}}
(W  \xi' + \bar{W} \bar{\xi'} ) & \nonumber \\
&- \frac{\sqrt{g^{55}}}{V} \delta(0) (4  W \bar{W} + V^{3/2} \bar{W} (\ov{\chi^{a}_R} \chi^{a}_L)) & \nonumber \\
&-D_\mu C_i D^\mu
\bar{C}^{i} - \frac{4}{V} \frac{\partial W_i}{\partial C_i}
\frac{\partial \bar{W}_i}{\partial \bar{C}^{i}}
 + \left ( \frac{W}{\sqrt{V}}
(\ov{\psi}_{L \mu} \gamma^{\mu \nu} \psi_{R\nu}) - \frac{1}{V^{2/3}} W (
\ov{\psi}_{L \mu} \gamma^\mu \lambda_L ) \right . & \nonumber \\  
& - \left . \frac{i}{V^{3/2}} \sqrt{g^{55}} W
(\ov{\psi}_{R5} \lambda_L) + h.c.
 ) \right )& \nonumber
\eeqa
where $i=1,2$ labels branes and $\epsilon_{1,2}=+1,-1$.\\
The  $Z_2$-odd 4d Majorana
supersymmetry
generator on the flipped brane is given by $
 \tilde{\epsilon} = \left (
\begin{array}{cccc}
{i\epsilon_{L}^{1}} \\
 {i\epsilon_{R}^{2}}
\end{array}
\right ) $ and analogous replacements hold for gravitini and hyperini.\\
This implies $\psi^{A}_\mu (2 \pi \rho + x^5) = - \psi^{A}_\mu (x^5)$
(and the same for hyperini) while bosonic fields remain periodic.
Obviously, there are no fermionic zero modes in the bulk, so supersymmetry is broken in that sector of the model. This has been explored in
\cite{Fabinger:2000jd} and more recently in \cite{gp,Gherghetta:2001kr}.
Bulk moduli still couple to both walls and participate
in the transmission of information between branes already at the classical level. The best example is again the odd scalar from the hypermultiplet, $\xi$.
If the expectation values of sources to which it couples on the walls
do not vanish, the $\xi$ and $\pa_5 \xi$ also assume nontrivial $x^5$
dependence, which creates operators breaking softly global supersymmetries
on the walls, similarly to what happens in the unflipped $(+,+)$ case.
{}From that case we know that on the unflipped wall the even components of
$\delta_{susy} \psi_5$ and $\delta_{susy} \lambda$ receive
nonhomogeneous contributions in their supersymmetry transformations.
What happens on the $\alpha=-1$ wall? Let us inspect the transformation law
of the odd part of the hyperini
\beq
\delta_- \lambda^{1}_R = ...-\frac{i}{2 \sqrt{2} V} \pa_5 (V + i \sigma)
\epsilon^{1}_R - \frac{i}{\sqrt{2 V}} \pa_5 \xi \epsilon^{2}_R .
\eeq
The generator $\epsilon^{1}_R$ is even, so it doesn't enter the wall.
The coefficent of the odd generator which generates supersymmetry on the wall
is multiplied by the parameter $\pa_5 \xi$, the same which induces
susy-breaking terms on the even wall. Hence, indeed, the walls talk to each other already at the level of the classical vacuum through the messenger $\xi$.
This communication obviously includes creation of supersymmetry breaking
terms on both walls, and these terms are sensitive to mass scales from the opposite wall, as in the $(+,+)$ models. The detailed analysis of this case will be given elsewhere, here we just point out the full analogy to the unflipped case in the  necessity  of arranging correllations between terms located at different walls in designing a phenomenologically relevant vacuum.\\

The above examples have illustrated the intimate interplay between the gravitational background and
the gauge sector physics in the case of flipped and unflipped locally supersymmetric brane world models. The backreaction of the gauge sector on the geometry is explicit, and shows that both sectors need to be tuned against each other to create a phenomenogically relevant vacuum.
In fact, generic vacua for generic values of the parameters in the Lagrangian are likely to be
cosmological ones, with time dependent geometry of the orbifold and physics on the brane.
The example of such a solution in models discussed here has been given in \cite{bfl}.

\section{Quiver models and deconstructed dimensions}

So far we were discussing the issues pertaining to a nondecoupling of gravity in brane world models.
However, there exist models where the decoupling of gravity from extra dimensions is natural,
and can be arranged by standard methods known from four-dimensional field theory.
Such are the models of deconstructed dimensions, \cite{ARCOGE_dec},\cite{HIPOWA},      
where gravity is always four-dimensional,
and extra dimensions are fictitious and fully contained within the 4d gauge sectors.
The example of such a situation  is provided by quivers with the custodial
supersymmetry \cite{bflp}.
The UV properties of these  models are  better than these of a generic non-supersymmetric model, and
a separation between the vevs on the gauge sector and the cut-off scale, which may be taken to be
the 4d Planck scale can be arranged.

\subsection{Orbifolding and supersymmetry breaking}
Consider the type IIB string theory with a  stack of n coinciding D3 branes.
It is well known that the gauge bosons and fermions living on the worldvolume of the D branes
form a 4d $N=4$ supersymmetric Yang-Mills model with  gauge group $U(n)$.
The six transverse dimensions form, from the point of view of the 4d theory living on the
worldvolume, six extra nongravitational dimensions.
One can obtain a theory with fewer supersymmetries than
$N=4$ $U(n)$ by dividing the extra dimensions by a discrete group $Z_\Gamma$ and
embedding this orbifold group into the gauge group $U(n\Gamma)$.
The resulting theory is called a quiver theory. We will focus on
non-supersymmetric quiver theories.
They are obtained by
retaining  in the spectrum only the fields which  are invariant under the combined
geometric and gauge actions of $Z_\Gamma$. Their interactions are consistently truncated   to yield
a smaller daughter gauge theory. The truncation process breaks the gauge group and
some (or all) supersymmetries. The gauge symmetry breaking is dictated by the embedding of the generator of
$Z_\Gamma$ into $U(n\Gamma)$. The matrix $\gamma$ that represents the gauge action of $Z_\Gamma$ is chosen to be of  the form of a direct sum of $\Gamma$ unit matrices of dimensions $n
\times n$, , each multiplied respectively
by $\omega^i$ with $\omega=e^{ \frac{2 \pi}{\Gamma} i}$.
Then the invariant components of the gauge fields
fulfill the condition
\beq
A = \gamma A \gamma^{-1}
\eeq
where $A$ is a matrix in the adjoint representation  of $U(n\Gamma)$.
This leaves invariant  the subgroup $U(n)^\Gamma$.
There are
four generations of Weyl fermions, each in the adjoint of $U(n\Gamma)$, whose invariant components
must obey the condition
\beq
\psi^i = \omega^{a_i} \gamma \psi^i \gamma^{-1}
\eeq
where $i=1,..,4$ and
\begin{equation}
a_1+a_2+ a_3 +a_4 =0.
\end{equation}
The invariant fermions transform in the
bifundamental representations of the broken gauge group $({\bf n}_l, \bar{\bf n}_{l +a_i})$
where $l$ numbers blocks of the original $n\Gamma \times n\Gamma $ matrices.
Furthermore, one obtains
three generations of complex bosons $\phi^i$, $i=1,2,3$, in the adjoint
of $U(K)$, whose invariant components fullfil the condition
\beq
\phi^i = \omega^{\tilde{a}_i} \gamma \phi^i \gamma^{-1}.
\eeq
The invariant scalars transform as $({\bf n}_l, \bar{\bf n}_{l+ \tilde{a}_i})$
under the broken gauge group.
The truncated fields have a block structure in the $U(n\Gamma)$ mother gauge group
\begin{equation}
\phi^i_{lp}=\phi_l^i\delta_{p,l+\tilde a_i},\ \psi_{lp}^i=\psi_{l}^i\delta_{p,l+a_i}
\end{equation}
Supersymmetry is preserved when  the group $Z_{\Gamma}$ is embedded in $SU(3)$
\begin{equation}
\tilde a_1+ \tilde a_2 +\tilde a_3=0
\end{equation}
In that case $a_4=0$ and at least one of the fermions can be paired with the gauge bosons, i.e. becoming a gaugino of $N=1$ supersymmetry.
We focus on the non-supersymmetric case $a_4\ne 0$.

Let us  move a stack of $n$  $D3$ branes from the origin.
{}From the field theory point of view, moving the stacks of $n$ $D3$
branes from the origin is equivalent to going to the Higgs  branch of
the theory where all the off-diagonal scalars with $\tilde a_i\ne 0$
take a vev
\begin{equation}
\phi^i_l=v^i1_{n\times n}
\end{equation}
Due to the
$Z_\Gamma$ action the stacks  have $\Gamma$ copies around the fixed point.
The gauge group is broken to the diagonal
subgroup $U(n)_D$. This is the deconstructed phase.

We will be interested in the one-loop divergences of the
non-supersymmetric quivers.
The vanishing
of the quadratic   divergences in models based upon the low energy dynamics
of branes in string theory is a general phenomenon. It
turns out to be an extension to the case of type II string theory
with D branes of misaligned supersymmetry.
Let us consider a model where the low energy fields live on a
$p$-brane in a configuration where many possible branes coincide.
Now assume that the background geometry is a solution
of the string equations with no closed string tachyons.
Let us first consider   that all the low energy fields vanish  so that
all the branes coincide. The open string mass spectrum comes from the
oscillators $M^2_0$.
When considering the case where  the  vev of  some of the low energy fields living on the
brane does not vanish, i.e. some of the brane have been moved,
the mass spectrum of open strings is shifted corresponding to the
minimal length of open strings between the branes.
Consider now the string amplitudes between any two of the displaced
branes
\begin{equation}
\hbox{Str}
\int_0^{\infty}\frac{dt}{t^{1+\frac{p+1}{2}}}e^{-{2\pi \alpha 'tM^2}}e^{-2\pi\alpha
'tM^2_O}
\end{equation}
where $M^2$ is the mass matrix of the low energy fields.
Open-closed duality relates this amplitude to
\begin{equation}
\int_{0}^{\infty} dl l^{\frac{p-9}{2}}<Dp'\vert e^{-\pi \alpha'M^2/l-\alpha 'lM_C^2}\vert Dp>
\end{equation}
where $l=1/2t$.
As $l\to\infty$ we find that
$
\hbox{Str}(M_T^{2k})=0,\ k=0\dots 3
$
where $M_T$ is the total mass matrix of all the open string states.
Now we consider  the decoupling limit $l_s \to 0$
sending all the stringy modes to infinity while preserving the
vev $v$ of the brane fields.
The low energy field theory on the stack of $D3$ branes is obtained
after decoupling gravity.
In the decoupling limit the mass matrix splits in two blocks
$
M_T=M\oplus M_O
$
acting on decoupled states.
We conclude that
\begin{equation}
\hbox{Str}{M^{2k}}=0,\ k=0\dots 3
\end{equation}
This is the vanishing of the supertraces corresponding to the
breaking of the low energy field theory by small (compared to the
string scale) vevs.
Of course this result is only valid when no closed string tachyon
propagate between the branes.


\noindent As soon as $a_4\ne 0$ there are closed string tachyons for non-supersymmetric quiver theories.
The twisted tachyons do not
intervene in the string amplitude when the branes are off the centre
of the orbifold. Indeed the boundary states of the branes are coherent
closed
string states satisfying
$
X^i\vert Dp>= x^i\vert Dp>
$
where $x^i$ is the location of the brane. If the twisted states couple
to the brane we must have
$
\theta x^i \equiv x^i
$
i.e. the brane is at a fixed point.
So we obtain the vanishing of the quadratic divergences in
the deconstructed phase.
More can be said here in the deconstructed phase.

The low energy degrees of freedom  come from the open strings
corresponding to the invariant states in the string spectrum
The action of  $Z_\Gamma$ leads to a truncation of the spectrum as only invariant
states are kept.
The orbifold $Z_\Gamma$ acts on the Chan-Paton indices by permutation implying that the action of $Z_\Gamma$ on states is
\begin{equation}
\theta \vert lp>=\omega^a \vert l+1, p+1>
\end{equation}
where $\theta$ is the generator of $Z_\Gamma$, $\vert lp>$ is a state
(either boson or fermion) with shift $a$ and the Chan-Paton indices $(lp)$ label the branes on which the open
strings end. Here each label $l$ corresponds to a stack of $n$ $D3$ branes and the
strings
are connected  to the mirror images of this stack.
The construction of invariant states follows
\begin{equation}
\vert lp>_{orb}=\sum_{k=0}^{\Gamma-1} \omega^{ka}\vert l+k,p+k>
\end{equation}
Notice that there are $\Gamma$ invariant states for each species.
When the stack of $n$ $D3$ branes is at the origin the length of the
open strings vanishes and the associated masses to the invariant state
is zero.
This gives rise to the low energy fields that we have discussed
previously.
Consider the $N=4$ mother theory written in terms of $N=1$ chiral
multiplets.
This is obtained by breaking the R-symmetry group from $SU(4)$ to $SU(3)\otimes Z_{\Gamma}$
where the three complex bosons $\phi^i$ are in the ${\bf 3}$ of $SU(3)$ and we
decompose the Weyl fermions as ${\bf 4}={\bf 3}+{\bf 1}$. The four spinors are distinguished by
their $Z_{\Gamma}$ charges which are respectively $a_i,\ i=1\dots 3$ for the ${\bf 3}$ and
$a_4$ for the singlet.
The $N=4$ fields can be arranged into $N=1$ supermultiplets $(\phi^i,\psi^i)$ and $(A_\mu, \psi^4)$.
In the orbifold theory, this $N=1$ invariance generated by a space-time supersymmetry generator $Q$ is broken when $a_4\ne 0$ as can be seen from the gauge
numbers of the fermions and bosons. Nevertheless the spectrum contains equal number of fermions and bosons, and these are paired up in a certain way.
To see this denote any  scalar by
\begin{equation}
\vert \phi >_{orb}(lp) =\sum_{k=0}^{\Gamma-1} \omega^{k\tilde a}\vert \phi > (l+k,p+k)
\end{equation}
and its associated fermion state  by
\begin{equation}
\vert \psi >_{orb}(lp) =\sum_{k=0}^{\Gamma-1} \omega^{k a}\vert \psi > (l+k,p+k)
\end{equation}
where $\vert \psi >(lp)$ and $\vert \phi >(lp)$ are superpartners under the action of $Q$.
Let us define the twisted supersymmetry operator
\begin{equation}
\tilde Q=\gamma^R Q\gamma^{-R}
\end{equation}
where $R$ is the $Z_\Gamma$ charge.
The action of $\tilde Q$ on the orbifold states is
\begin{equation}
(\tilde Q\vert \phi >_{orb})(lp) = \omega^{-la_4}\vert \psi >_{orb}(lp).
\end{equation}
This implies that the string states, i.e. the physical fields in the low energy limit
are classified into twisted supersymmetry multiplets. However, this is a kinematical statement,
and one needs to examine masses and interactions to draw stronger conclusions.

\subsection{Orbifolding of the field-theoretical Lagrangian and custodial supersymmetry}

To prove more about non-supersymmetric quivers
it is very useful to study the effective Lagrangian of such theories.
Inserting the block decomposition  into the ${\cal N} =4$ lagrangian we find the daughter theory lagrangian
\bea &
\label{eq:daughter}
\cl=
Tr \left \{
-{1 \over 2} F_{\mu \nu,p} F_{\mu\nu,p} + i \ov{\lambda_p} \gamma^\mu D_\mu \lambda_p
+ 2 D_\mu \phi_{i,p}^\dagger D_\mu \phi_{i,p}  + i \ov{\psi_{i,p}} \gamma^\mu D_\mu \psi_{i,p} \right .
&\nn&
- g_0\left [2i\rt ( \ov{\psi_{i,p}} P_L \lambda_{p+a_i} \phi_{i,p}^\dagger -  \ov{\psi_i}  \phi_{i,p-a_4}^\dagger P_L \lambda_{p-a_4})
+ {\rm h.c.} \right]
&\nn&
-g_0\left [i\rt \epsilon_{ijk} (
\ov{\psi_{i,p}} P_L \psi_{j,p+a_i} \phi_{k,p-\tilde a_k}
 - \ov{\psi_{i,p}} \phi_{k,p+a_i} P_L \psi_{j,p-a_j} ) + {\rm h.c.}
\right ]
&\nn&
-g_0^2(\phi_{i,p} \phi_{i,p}^\dagger
- \phi_{i,p-\tilde a_i}^\dagger\phi_{i,p -\tilde a_i })
(\phi_{j,p} \phi_{j,p}^\dagger
- \phi_{j,p-\tilde a_j}^\dagger\phi_{j,p -\tilde a_j })
&\nn& \left.
+4 g_0^2 (\phi_{i,p} \phi_{j,p+ \tilde{a}_i}
 \phi_{i,p+ \tilde{a}_j}^\dagger  \phi_{j,p}^\dagger
- \phi_{i,p} \phi_{j,p+ \tilde{a}_i} \phi_{j,p+ \tilde{a}_i}^\dagger \phi_{i,p}^\dagger) \right \}.
&\nn&
\eea
where $\lambda\equiv \psi_4$.
The covariant derivative acting on scalars is
$D_\mu\phi_{i,p} = \pa_\mu \phi_{i,p} + ig_0A_p\phi_{i,p} -ig_0\phi_{i,p}A_{p+\ta_i}$.
It is then a tedious exercice to obtain the mass matrices
and compute the super-trace
\bea
\label{eq:str}
STr ({\cal M}^2) =
4g_0^2 \sum_k \sum_p \delta_{\ta_k,0} \left [
\left ( Tr (\phi_{k,p}^\dagger) Tr(\phi_{k,p+a_4})
+Tr(\phi_{k,p+a_4}^\dagger)Tr (\phi_{k,p})
-2 Tr (\phi_{k,p}^\dagger) Tr(\phi_{k,p}) \right)\right.
\nn + \left . \sum_i
 \left(
Tr (\phi_{k,p}^\dagger) Tr(\phi_{k,p+a_i})
+  Tr(\phi_{k,p+a_i} ^\dagger)Tr (\phi_{k,p})
-Tr (\phi_{k,p}^\dagger) Tr(\phi_{k,p+\ta_i})
-  Tr(\phi_{k,p+\ta_i} ^\dagger)Tr (\phi_{k,p}) \right )
\right].
&\nn&
\eea
One can check  that (\ref{eq:str}) vanishes identically if at least one of the following conditions is satisfied:
\bi
\item $a_4=0$ or $a_i=0$, that is when at least ${\cal N}=1$ supersymmetry is preserved by the orbifolding,
\item $\ta_1 \neq 0 $, $\ta_2 \neq 0$  $\ta_3 \neq 0$, that is when there are no scalars in adjoint representation of $U(n)$ group.
\ei
In the first case the vanishing of the supertrace is of course guaranteed by unbroken supersymmetry of the daughter theory.
Surprisingly, the absence of quadratic divergences can also occur if the daughter theory is  completely non-supersymmetric, the only condition being that all scalars are in bifundamental representations of the $U(n)^\Gamma$ gauge group.
This result is stronger than the result that we derived
previously from stringy arguments. Indeed it is valid for any
background value of the six scalar fields.

Let us now come back to the deconstructed case.
One can explicitly diagonalize the mass matrices. For instance
 the gauge bosons acquire mass terms:
\beq
\cl =  \sum_p \sum_{k=1}^3 g_0^2 v_k^2 (A_p^a - A_{p+\ta_k}^a)^2.
\eeq
(We have rewritten the gauge fields as $A = A^aT^a$ and evaluated the trace over generators. In the following we often omit the adjoint index $a$.) These mass terms are diagonalized by the following mode decomposition \footnote{The
 decomposition is given for odd $\Gamma$. For even $\Gamma$ the first sum goes to $\Gamma/2$ and the second to $\Gamma/2-1$.}:
\beq
\label{eq:gfmd}
A_p = \sqrt{2 \over \Gamma} \left (
\sum_{n=0}^{(\Gamma-1)/2} \eta_n \cos \left( {2 n \pi \over \Gamma}p\right)A^{(n)}
+\sum_{n=1}^{(\Gamma-1)/2}\sin \left( {2 n \pi \over \Gamma}p\right)\tilde A^{(n)} \right).
\eeq
where $\eta_0=1/\sqrt 2$ and $\eta_n=1,\ n\ne 0$.
Plugging in this decomposition we get:
\beq \label{spectr}
\cl =  {1\over 2}\sum_n\sum_k (m_k^{(n)})^2 (A^{(n)}A^{(n)} +\tilde A^{(n)}\tilde A^{(n)})\hspace{2cm} m_k^{(n)} \equiv 2\rt g_0 v_k \sin \left( {n \pi \over \Gamma}\ta_k\right),
\eeq
so that the $n$-th level gauge bosons have masses  $(m^{(n)})^2 = \sum_k m_k^2$.
Similar calculations can be done for the other fields resulting
in the
fact that  the spectrum is perfectly boson-fermion degenetate. We
already know that this degeneracy can be traced back to a
custodial supersymmetry. Let us now investigate it further.
We define the  vector superfields in the Wess-Zumino gauge as:
\beq
V^{(n)}(y,\theta) = {i \over 2}(\ov{\theta} \gamma_5 \gamma_\mu \theta) A^{(n)}
  -i  (\ov{\theta} \gamma_5 \theta) (\ov{\theta} \lambda^{(n)})
-{1 \over 4} (\ov{\theta} \gamma_5 \theta)^2 D^{(n)}.
\label{t1}
\eeq
Similarly we define chiral superfields:
\beq
\Phi_i^{(n)}(y,\theta) = X_i^{(n)} - \rt  (\ov{\theta} P_L \psi_i^{(n)})
+ F_i^{(n)}(\ov{\theta} P_L \theta).
\label{t2}
\eeq
Analogous expressions for the tilded fields hold.

First, we note that the self-couplings in  the zero-mode sector  are those of the ${\cal N}=4$ supersymmetric theory. Indeed,  the interactions of the zero-modes can be found by making in (\ref{eq:daughter})  the replacement $\phi_{i,p} \ra {1 \over \sqrt{\Gamma}} \phi_i^{(0)}$ (and similarly for fermion and gauge fields). Since all memory of the block indices is lost,
as a result we obtain  the ${\cal N}=4$ lagrangian with  gauge coupling $g = {g_0 \over \sqrt{\Gamma}}$. Second, we have already shown that the mass pattern in the deconstruction phase is supersymmetric. It turns out that the custodial supersymmetry has a much wider extent and all the terms quadratic in the heavy modes (including triple and quartic interactions with the zero-modes) match the structure of a globally supersymmetric theory! As an example we present a superfield lagrangian which reproduces the  Yukawa terms and the scalar potential of the daughter theory:
\bea
&\cl = \sum_n \sum_k {\rm Tr} \left [
4 g_0 v_k \sin \left({ n \pi \ta_k \over \Gamma}\right) \left (
 \tilde V^{(n)} \Phi_k^{(n)} -   V^{(n)} \tilde \Phi_k^{(n)} \right)
\right. &\nn&
+ 2 g  \cos \left({ n \pi \ta_k \over \Gamma}\right)
 \left ( [\Phi_k^{(0)\dagger},\Phi_k^{(n)}] V^{(n)}  +
[\Phi_k^{(0)\dagger},\tilde \Phi_k^{(n)}] \tilde V^{(n)} \right)
&\nn& \left .
+  2 g  \sin \left({ n \pi \ta_k \over \Gamma}\right)
 \left ( \{ \Phi_k^{(0)\dagger},\Phi_k^{(n)} \} \tilde V^{(n)}  -
\{\Phi_k^{(0)\dagger},\tilde \Phi_k^{(n)}\} V^{(n)} \right)
+{\rm h.c.}
\right ]_D
&\nn&
+ [W]_F  + [W^*]_F ,
\eea
where  the superpotential is:
\bea
&
W =  - i \sqrt{2} \sum_n \sum_{ijk} \epsilon_{ijk} {\rm Tr } \left [
4 g_0 v_k \sin \left({ n \pi \ta_k \over \Gamma}\right) \Phi_i^{(n)} \tilde \Phi_j^{(n)} \right. &\nn&
- g  \cos \left({ n \pi \ta_k \over \Gamma}\right)
 \left ( [\Phi_k^{(0)},\Phi_i^{(n)}] \Phi_j^{(n)}  +
[\Phi_k^{(0)},\tilde \Phi_i^{(n)}] \tilde \Phi_j^{(n)} \right)
&\nn& \left .
+  g  \sin \left({ n \pi \ta_k \over \Gamma}\right)
 \left ( \{ \Phi_k^{(0)},\Phi_i^{(n)} \} \tilde \Phi_j^{(n)}  -
\{\Phi_k^{(0)},\tilde \Phi_i^{(n)}\} \Phi_j^{(n)} \right)  \right ].
\eea
 Supersymmetry is explicitly violated by triple and quartic self-interactions of the heavy modes. Nevertheless, the presence of the custodial supersymmetry in the lagrangian is sufficient to ensure the vanishing  of  one-loop corrections to the zero-mode masses. A mass-splitting of the zero-mode multiplets can appear only at the two-loop level and we expect the supersymmetry breaking scale to be suppressed $M_{\rm SUSY} \ll v \ll \Lambda$.

\subsection{Theory space dimensions}
{}From the previous discussion we know that the daughter theory is the low-energy field theory of branes located at the fixed point of
an  orbifold. The low energy degrees of freedom on a brane are those combinations
of the open string states that are invariant
under the action of  $Z_\Gamma$.
When one
moves  a stack of $n$  $D3$ branes at a distance $d$ away from the
fixed point, due to the $Z_\Gamma$ symmetry there appear $\Gamma$ copies
of the stack, spaced symmetrically in the transverse directions around the fixed point, see 
\cite{ARCOKA}.
The custodial supersymmetry implies an extension
of the results of Arkani-Hamed et. al.  to nonsupersymmetric orbifoldings.
For instance, it was shown  that in the large $\Gamma$ limit, when the distances between images of the stack are much smaller than $d$, one can redefine the orbifold metric in such a way,
that consecutive boson-fermion degenerate mass
levels correspond to open strings winding around a circular direction of the transverse geometry.
 This geometric picture allows for the
straightforward computation of the massive string spectrum:
\beqa
&m_n^2= 4 \frac{d^2}{l_s^4}\sum_{i=1}^3\sin^2( \frac{n \pi \tilde a_i}{\Gamma}),&
\nonumber \eeqa
where $l_s$ is the string scale and the shifts $\tilde a_i$ represent the action of $Z_\Gamma$ on the three complex coordinates. When all vevs are equal, this is precisely the field theoretical spectrum
in the deconstruction phase of the nonsupersymmetric model.
In fact one can forget about the underlying stringy picture, and view the additional dimensions as fictitious, theory space, dimensions\footnote{Actually, each allowed set of shifts defines a closed subset of links in quiver diagrams, which can be interpreted as an internal dimension}.
The ladder of scales which appears in a deconstructed field theoretical quiver model is as follows.
 The first scale one encounters, taking the bottom-up direction in available energy,  is the fictitious compactification scale
$1/R_5=agv/\Gamma$, where  $a^2=\sum_i \tilde a_i^2$.
At this scale a seeming  fifth dimension opens up and one sees
the tower of Kaluza-Klein states with masses of order $1/R_5$. Hence above this scale
the theory looks five-dimensional.
Moreover the spectrum of massive states is determined by the custodial supersymmetry.
This picture holds up to the deconstruction scale $v$ where non-diagonal gauge bosons
become massless again. Above the deconstruction scale the theory
is explicitly four-dimensional, nonsupersymmetric and renormalizable.
Quadratic divergences are absent at the one-loop level.
Also at one-loop the deconstruction scale is a flat direction of this four dimensional theory, hence it stays decoupled
from any UV cut-off scale, including the 4d Planck scale.
Moreover the compactification scale $1/R_5$ can be arbitrarily smaller
than the deconstruction scale and it is determined by the discrete parameter which is the
order $\Gamma$ of the orbifold group $Z_{\Gamma}$.

\section{Summary}

In this paper we have discussed the interplay between gravity/moduli  and gauge/matter sectors
in creating physically relevant vacua in higher-dimensional brane worlds. We started with
five-dimensional brane-bulk supergravities, with flipped and unflipped boundary conditions, constructed and discussed in \cite{flp,flp2,flp3}, \cite{bfl}. We have demonstrated that in both, flipped and unflipped cases,
adding nontrivial gauge/matter sectors on the branes changes geometry of the four-dimensional
sections and affects stabilization of the orbifold. To achieve vacua with hierarchically broken supersymmetry, static orbifold and nearly vanishing 4d cosmological constant one needs a tuning
involving all sectors of the brane-bulk Lagrangian. Thus the physics of visible and gravity/moduli
sectors cannot be treated separately in such models, and the need for a tuning translates into the issue of stability of mass scales.
A possibility  for  a natural separation between the gauge and gravity sectors, achieved by field-theoretical methods known from four-dimensional theories,  appears in models
with deconstructed dimensions. As an  example of deconstruction
we have discussed quiver theories which result from a
nonsupersymmetric orbifolding of the ${\cal N}=4$ $U(K)$ gauge theories.
In a generic situation these non-supersymmetric models  exhibit an
improved UV behaviour - the quadratically divergent contributions to the effective potential vanish at the one-loop level. If the gauge group resulting from orbifolding becomes broken down to the diagonal
subgroup by universal vevs, then the resulting low-energy theory exhibits custodial supersymmetry and
theory space extra dimensions.
The hierarchy $v \ll M_{cut-off}$ is protected at the one-loop level, and
 at one-loop universal vevs remain a flat direction and  zero-mode multiplets do not suffer
from a mass splitting. Of course, the situation becomes even better in $N=1$ supersymmetric orbifoldings.
The deconstructed extra dimensions are fictitious, and belong to a renormalizable 4d gauge model.
On the other hand gravity is four-dimensional at all scales. Hence, while retaining at low energies
signatures of extra dimensions, these models simplify the physics of the gauge sector/gravity
interface.\\
\vskip0.7cm

{\bf Acknowledgments:}
The authors would like to thank Stefan Pokorski and Adam Falkowski for collaboration on the issues discussed in this paper.\\
\vskip0.1cm
The work of P.B. and Z.L. has been supported by RTN programme HPRN-CT-2000-00152,
Polonium 2002,
and by the Polish Committee for Scientific Research grant
5~P03B~119 20 (2001-2002).

\end{document}